\documentclass[acmsmall]{acmart}

\AtBeginDocument{%
  \providecommand\BibTeX{{%
    \normalfont B\kern-0.5em{\scshape i\kern-0.25em b}\kern-0.8em\TeX}}}

\setcopyright{rightsretained}
\copyrightyear{2024}
\acmYear{2024}
\acmDOI{XXXXXXX.XXXXXXX}

\acmConference[ISS '24]{Interactive Surfaces and Spaces Conference}{Oct,
  2024}{Vancouver, BC}
\acmISBN{978-1-4503-XXXX-X/18/06}

\begin{document}

\title[Experimental Analysis of Freehand Multi-Object Selection Techniques in VR HMDs]{Experimental Analysis of Freehand Multi-Object Selection Techniques in Virtual Reality Head-Mounted Displays}

\author{Rongkai Shi}
\orcid{0000-0001-8845-6034}
\authornote{Both authors contributed to this research when they were students at the University of Liverpool and the Xi'an Jiaotong-Liverpool University.}
\affiliation{%
  \institution{The Hong Kong University of Science and Technology (Guangzhou)}
  \city{Guangzhou}
  \country{China}}
\email{rongkaishi@hkust-gz.edu.cn}

\author{Yushi Wei}
\orcid{0000-0002-6003-0557}
\authornotemark[1]
\affiliation{%
  \institution{The Hong Kong University of Science and Technology (Guangzhou)}
  \city{Guangzhou}
  \country{China}}
\email{ywei662@hkust-gz.edu.cn}

\author{Xuning Hu}
\orcid{0009-0009-1305-2081}
\affiliation{%
  \institution{Xi'an Jiaotong-Liverpool University}
  \city{Suzhou}
  \country{China}}
\email{xuning.hu22@student.xjtlu.edu.cn}

\author{Yu Liu}
\orcid{0000-0003-0226-1311}
\affiliation{%
  \institution{Xi'an Jiaotong-Liverpool University}
  \city{Suzhou}
  \country{China}}
\email{yu.liu@xjtlu.edu.cn}

\author{Yong Yue}
\orcid{0000-0001-7695-4538}
\affiliation{%
  \institution{Xi'an Jiaotong-Liverpool University}
  \city{Suzhou}
  \country{China}}
\email{yong.yue@xjtlu.edu.cn}

\author{Lingyun Yu}
\orcid{0000-0002-3152-2587}
\affiliation{%
  \institution{Xi'an Jiaotong-Liverpool University}
  \city{Suzhou}
  \country{China}}
\email{lingyun.yu@xjtlu.edu.cn}

\author{Hai-Ning Liang}
\orcid{0000-0003-3600-8955}
\authornote{Corresponding author.}
\affiliation{%
  \institution{The Hong Kong University of Science and Technology (Guangzhou)}
  \city{Guangzhou}
  \country{China}}
\email{hainingliang@hkust-gz.edu.cn}

\renewcommand{\shortauthors}{Shi, et al.}

\begin{abstract}
Object selection is essential in virtual reality (VR) head-mounted displays (HMDs). Prior work mainly focuses on enhancing and evaluating techniques for selecting a single object in VR, leaving a gap in the techniques for multi-object selection, a more complex but common selection scenario. To enable multi-object selection, the interaction technique should support group selection in addition to the default pointing selection mode for acquiring a single target. This composite interaction could be particularly challenging when using freehand gestural input. In this work, we present an empirical comparison of six freehand techniques, which are comprised of three mode-switching gestures (Finger Segment, Multi-Finger, and Wrist Orientation) and two group selection techniques (Cone-casting Selection and Crossing Selection) derived from prior work. Our results demonstrate the performance, user experience, and preference of each technique. The findings derive three design implications that can guide the design of freehand techniques for multi-object selection in VR HMDs. 
\end{abstract}



\received{20 February 2007}
\received[revised]{12 March 2009}
\received[accepted]{5 June 2009}

\maketitle

\section{Introduction} \label{sect:intro}
Object selection (or target acquisition) is a fundamental interaction in virtual reality (VR). It is typically the initial step to complete other canonical manipulation tasks, including positioning, rotation, and scaling, and is indispensable for composite workflows~\cite{3DUIBook2017, Argelaguet2013Survey}. Numerous prior work has proposed interaction techniques to improve object selection and enable their use in complex VR scenarios~\cite{Argelaguet2013Survey, 3DUIBook2017, Bergstrom2021How}, such as selecting a target that is small, out-of-reach, (e.g.,~\cite{Poupyrev1996Go-Go, Mendes2017Design}), or occluded by others (e.g.,~\cite{Yu2020Fully, Maslych2023Towards}). However, most of these interaction techniques share the same goal---to acquire a single target, leaving fewer discussions on multi-object selection, which is common in VR applications involving large amounts of selectable objects, such as astronomical data exploration~\cite{Zhao2024MeTACAST} and 3D modeling design~\cite{Xia2018Spacetime, Shi2022Group}.

When a user has the same operation intention for multiple objects, selecting them first and manipulating them as a group would be less time-consuming and tedious than repetitively working with each object, especially when there are many intended targets or the manipulation requires high precision~\cite{Lucas2005Design, Shi2022Group, Wu2023Point}. Furthermore, performing a group selection could be effortless if the intended targets are located in a particular area~\cite{Shi2023Exploring}. Compared to single-object selection, multi-object selection can be relatively more challenging because it requires additional iterations for refining the selection, like deselecting unwanted objects or appending other non-selected targets, and in the meantime, extra effort for holding the selection state for selected targets while completing the refinement. Moreover, as a special case of single-object selection, enabling multi-object selection introduces an additional mode to the interaction scenario, which may make users more prone to make mistakes when performing or transitioning between the two types of selection. 

The multi-object selection techniques proposed in prior work are largely based on handheld devices, like pen-tablet combinations~\cite{Lucas2005Design} or controllers~\cite{Xia2018Spacetime, Wu2023Point}, while limited supports freehand gestural input, which is becoming popular. Freehand gestural input is controller-free and has the potential to facilitate effective, natural, immersive interaction and communication~\cite{Buckingham2021Hand}. It is supported by current VR head-mounted displays (HMDs) without extra devices and has become an alternative to handheld controllers. However, there are several challenges for freehand gestural interaction, such as relatively high learning costs, limited available delimiters/triggers, lack of tactile feedback, and absence of critical clues for interaction~\cite{Norman2010Natural, Ren2013Selection}. Moreover, prior work mainly focused on near-field virtual-hand-based selection for multiple objects, leaving a gap in far-field virtual-pointing-based selection, which is also common and essential in the immersive VR environment to overcome physical constraints~\cite{Argelaguet2013Survey, Mine1995Virtual}. Given these challenges from both the selection task and the gestural input, enabling precise and efficient freehand selection for multiple objects in addition to pointing selection of a single object requires careful designs.

Thus, our research goal is to \textbf{design and evaluate freehand techniques for multi-object selection in VR HMDs}. To achieve this goal, we first frame three design goals in the context of related work to guide the interaction design. We then analyzed the interaction process of selecting multiple objects and derived three pivotal actions for this process: default single-object selection, group-based multi-object selection, and mode switching. The thumb-to-index pinch gesture, the most widely adopted hand gesture for freehand pointing selection, has been chosen for single-object selection. Building upon this, six potential techniques were proposed and selected for evaluation. These techniques are the combinations of three mode-switching gestures (Finger Segment, Multi-Finger, and Wrist Orientation) and two group selection techniques (Cone-casting Selection and Crossing Selection) derived from prior work. They were compared empirically via a user study with eighteen participants in randomized scenarios. We found Crossing Selection outperformed Cone-casting Selection while the latter was not disliked by participants. The three mode-switching gestures led to similar performance and user experience. Participants tended to like Multi-Finger and to dislike Wrist Orientation. Our findings are useful for the future design of freehand interaction techniques for multiple objects in VR environments. 

This work makes the following three main contributions: 
\begin{itemize}
    \item We articulate three design goals for freehand multi-object selection in VR based on a synthesis of previous work (Section~\ref{sect:DG}).
    \item Based on the design goals, we propose six freehand multi-object selection techniques that combine three mode-switching gestures and two group selection techniques (Section~\ref{sect:design}).
    \item We empirically compare the six techniques via a user study and derive insights for future design and development of freehand techniques for multi-object selection in VR (Sections \ref{sect:exp}, \ref{sect:results}, and \ref{sect:discussion}). 
\end{itemize}

\section{Design Goals and Related Work} \label{sect:DG}
We identified three design goals for a freehand interaction technique supporting precise and efficient multi-object selection in VR. In this section, we frame these design goals in the context of related work. 

\subsection{Design Goal 1: Build upon General Interaction Metaphors for Freehand Object Selection} \label{sect:DG:selection}
There are two major interaction metaphors for object selection in VR environments~\cite{Argelaguet2013Survey}---\textit{virtual hand}~\cite{Mine1997Moving} and \textit{virtual pointing}~\cite{Mine1995Virtual}. The interaction provided by virtual hand techniques is widely considered natural and intuitive because users interact with virtual objects in a similar way as they do in the real world. However, this mapping also limits the use of the original virtual hand technique because the interaction only happens in users' reachable areas. Two common solutions are identified for letting users select out-of-reach objects, 1) using a technique that sends the virtual hand out and controls it remotely, such as Go-Go~\cite{Poupyrev1996Go-Go}, and its extensions (e.g.,~\cite{Frees2005Precise, Wilkes2008Advantages}); and 2) using virtual pointing techniques. The most common virtual pointing technique is ray-casting~\cite{Mine1995Virtual}. With a ray-casting technique, the user casts a ray emitted from the controller or the hand, allowing the user to select an object at a distance. However, its performance also suffers from difficulties in selecting small objects and hand jitter issues when triggering the selection (i.e., the Heisenberg effect~\cite{Bowman2001Using}). The pointing accuracy has been improved via correction models from both the human side~\cite{Dalsgaard2021Modeling} and the system side~\cite{Mayer2018Effect}. To summarize, the two major interaction metaphors have their advantages and disadvantages. 

Currently, top VR headset and hand-tracking accessory companies suggest a combined use of virtual hand and virtual pointing techniques for their hand-tracking solutions (see, for example, Meta Quest\footnote{\url{https://www.meta.com/en-gb/help/quest/articles/headsets-and-accessories/controllers-and-hand-tracking/hand-tracking-quest-2/}}, HTC VIVE\footnote{\url{https://www.vive.com/au/support/focus3/category_howto/hand-tracking.html}}, and Ultraleap\footnote{\url{https://docs.ultraleap.com/xr-guidelines/Getting\%20started/design-principles.html}}). A user can tap on a target positioning within their reach for direct selection. On the other hand, the user can point the ray to a target and then pinch their thumb and index finger together for distant selection. This work considers this general metaphor group the foundation of enhanced technique, though there are also other hand gestures feasible for object selection (e.g., bending the thumb for selection~\cite{Ishii2017FistPointer}). We discuss the design of selection techniques in detail in Section~\ref{sect:design}.

\subsection{Design Goal 2: Facilitate Effective Multi-Object Selection} \label{sect:DG:MOS}
There are two main approaches to selecting multiple objects: selecting objects serially or selecting by group; that is, selecting one object versus one or more objects per selection operation~\cite{Wu2023Point, Lucas2005Design, 3DUIBook2017}. 

The single-object selection techniques can be used for serial selection. Prior empirical results have shown that serial selection is necessary for certain scenarios because adding or removing an object that is more challenging using group selection metaphors (like a distractor surrounded by targets) is unavoidable~\cite{Lucas2005Design, Wu2023Point}. We identified two ways to provide the serial selection mode. In the daily use of a desktop computer, users can activate the serial selection mode by pressing the shift key. While prior work focusing on multi-object selection in VR did not distinguish between these two operations---a new selection will not change the selection states of others~\cite{Lucas2005Design, Wu2023Point}. This work followed the way of a desktop computer to enrich VR interactions (single-object selection and serial selection as two operations), which is still underexplored. 

Four existing metaphors are possibly suitable for group selection. One such metaphor is goal crossing, which has been introduced to various selection scenarios in HMDs~\cite{Tu2019Crossing, Uzor2021Exploration, Huang2019Review}. Simply put, it works like a brush and can select an object by interacting with its boundary. With the selection activated and maintained, users can select multiple objects. Notably, the ray-casting crossing has been verified as a feasible complement to the ray-casting pointing~\cite{Tu2019Crossing}. In the desktop and tablet interfaces, users can `click and drag' to complete a rectangle selection for selecting multiple targets. Rectangle selection has been adapted to the 3D world by Shi et al.~\cite{Shi2023Exploring}, who explored gaze-assisted and hand-based region selection methods in AR HMD. In their hand-only region selection method, users can pinch and drag to formulate a rectangular region, which can potentially be used to select the objects cast by this region. Except for a 2D region, prior studies have also explored selecting objects via a 3D volume in VR environments. For example, Lucas~\cite{Lucas2005Design} allowed multi-object selection by creating a cuboid region via tablet and stylus input. Wu et al.~\cite{Wu2023Point} adapted this approach to VR controllers, enabling group selection/deselection in the near-field with the simulated virtual hands. Another 3D volume proposed in the literature is a spherical container, see for example, Poros~\cite{Pohl2021Poros}, SpaceTime~\cite{Xia2018Spacetime}, and BodyOn~\cite{Yu2022Blending}. These works also focused on the interaction within arm's reach. For far-field techniques, cone-casting, which replaced the ray with a cone or a spotlight, may be suitable for group selection. It is a widely investigated pointing technique for assisting target selection in dense environments or small objects at a distance (e.g.,~\cite{Mendes2017Design, Shi2023Exploration, Yu2020Fully}). Cone-casting makes the single-target acquisition easier and more precise by enabling users to select the target from a small group of objects, which is pre-selected via the cone and rearranged to a structured layout. Its first step is an obvious multi-object selection process but yet to be examined. 

As seen above, the prior work has provided a few interaction approaches and metaphors with good potential for multi-object selection. This work proposes effective freehand techniques for multi-object selection based on their exploration.

\subsection{Design Goal 3: Enable Rapid and Seamless Mode Switching} \label{sect:DG:MS}
Mode switching is the transition between different modes, which allows users to achieve different outcomes with the same input~\cite{Raskin2000Humane}. Researchers have highlighted the importance of designing suitable mode-switching methods for freehand selection and manipulation~\cite{Shi2023Exploration, Chen2020Disambiguation, Yu2022Blending} because the same action is inevitably needed for multiple purposes. Furthermore, rapid and seamless mode switching is necessary and particularly important for freehand multi-object selection, given the challenges we identified in Section~\ref{sect:intro}. 

There are three common mode-switching mechanisms. First, a mode can be sustained by the system (i.e., \textit{system-maintained}~\cite{Sellen1992Prevention}). The system persistently activates the selected mode until the user switches to others (e.g., the Caps Lock key). It is useful when several operations are pending in the selected mode. Second, a mode can be maintained by users (\textit{user-maintained}~\cite{Sellen1992Prevention} or \textit{quasi modes}~\cite{Raskin2000Humane, Hinckley2006Springboard}), which requires users to `hold' the switching action as long as the mode is needed. A user-maintained mechanism can help reduce mode errors~\cite{Sellen1992Prevention, Raskin2000Humane} and is suitable for temporary use~\cite{Hinckley2006Springboard}. Third, a mode is manually activated for one use only and then automatically returns to the previous mode (\textit{Once}~\cite{Hinckley2006Springboard}), which can be ideal for certain cases, such as typing the first letter of a sentence. This work mainly focused on the user-maintained mechanism as we consider multi-object selection a special and temporarily performed task in addition to the regular single-object selection. User-maintained mode-switching methods are also widely studied for different VR tasks and reported to be usable in prior work~\cite{Wan2024Design, Shi2021Exploring, Song2022Efficient, Surale2019Experimental}. 

In freehand interaction, hand postures can be used to distinguish between different modes. When both hands are available, the dominant hand is assigned to the primary task, leaving the non-dominant hand to control the mode naturally. Using non-dominant hand posture for mode switching has been proven to be efficient and accurate~\cite{Smith2020Evaluating, Park2019HandPoseMenu, Surale2019Experimental}. For one-hand cases, Surale et al.~\cite{Surale2019Experimental} compared hand gestures empirically and suggested subtle dominant hand postures, such as rotating the wrist or using the middle finger to distinguish from the default thumb-to-index pinch. Similarly, Song et al.~\cite{Song2022Efficient} enabled efficient keyboard switching for freehand text entry by rotating the wrist or extending the middle finger while performing the finger-touch selection. On the other hand, Yu et al.~\cite{Yu2022Blending} used tapping the thumb on different fingertips or finger segments to achieve various manipulations with three levels of control-display ratios. These small adjustments to the hand gesture for mode switching were considered in this work (see Section~\ref{sect:design}). Introducing a secondary input modality for mode switching to support the interaction is a huge branch of related research (e.g., voice~\cite{Smith2019Experimental, Chen2020Disambiguation}, eye gaze~\cite{Pfeuffer2017Gaze, Pfeuffer2020Empirical}, or head movement~\cite{Shi2021Exploring, Chen2020Disambiguation, Wan2024Hands-free}), but it is out of the scope of this paper. 

\section{Design of Techniques} \label{sect:design}
We aim to investigate user performance and experience of possible freehand techniques for multi-object selection in VR HMDs. In this section, we first propose a set of considerations to formulate the problem (Section~\ref{sect:design:considerations}), followed by an analysis of the interaction process (Section~\ref{sect:design:workflow}). Based on this, we present 12 potential technique combos consisting of 3 mode-switching techniques and 4 group selection techniques, and 6 were selected for evaluation (Sections~\ref{sect:design:techniques} and \ref{sect:design:summary}).

\subsection{Considerations} \label{sect:design:considerations}
To begin with, we describe several considerations that narrow down the problem space of this work. First, our domain selection metaphors are based on an egocentric point of view (first-person view), which is the most common for immersive environments and receives much attention from prior work on object selection. The interaction design of multi-object selection is built upon virtual-pointing-based selection for acquiring a single object. Second, we focus on selection-intensive scenarios that do not involve navigation. Though navigate-to-select approaches are interesting and important, they may generate potential issues beyond selection~\cite{Argelaguet2013Survey}, which are outside this paper's scope. Third, we do not consider intelligent grouping facilitated by the system (e.g.,~\cite{Zhao2024MeTACAST}). In other words, the selection can only be driven and completed by users' intentions. Finally, we define \textit{freehand} as input performed entirely by a hand gesture or movement. In this work, we utilized the self-built hand-tracking modules on VR HMDs to track freehand interaction, which we believe is low-cost and will be commonly available on future HMDs. Even so, we admitted this solution was imperfect, and as such, we focused on and ensured that the proposed interaction was doable for current commercial HMDs.

\subsection{Interaction Process} \label{sect:design:workflow}
We view multi-object selection as an auxiliary case of single-object selection, not a standalone and individual task. Thus, single-object selection is still the default mode for interaction, while users can switch to the multi-object selection mode seamlessly and have a fluid workflow for the whole process. On the other hand, as shown in Section~\ref{sect:DG:MOS}, prior work has shown empirical evidence suggesting serial selection for randomized target layout~\cite{Wu2023Point}. Thus, both default single-object and serial selection were included in the multi-object selection workflow. For simplicity and clarity, we call them \textbf{Default Selection} and \textbf{Serial Selection} to describe the object-of-interest cases in which only one object can be selected (like click on a PC mouse) and multiple objects can be selected (`shift + click'), respectively. Additionally, we use the term \textbf{Group Selection} to describe the action of acquiring several targets at a time. Users can switch between them via \textbf{Mode-Switching} methods. When a user performs serial selection or group selection to objects that have been selected, the selection states of those objects would be canceled (i.e., deselection). 

However, in our pilot tests, we faced similar hand-tracking issues as reported in prior work (e.g., \cite{Shi2023Exploration}). We found it difficult for current VR HMDs to track accurately the micro-gestures for switching between three modes.\footnote{The tested HMDs included Quest 2, 3, and Pro.} We were also concerned about the high learning cost of gestures for people new to VR and mid-air interaction~\cite{Norman2010Natural, Ren2013Selection}. Thus, we improved the interaction process by \textbf{making the serial selection an implicit part of group selection}, which does not require an explicit mode switch to distinguish between them. This approach has a lower hand-tracking requirement for HMDs but also eases users' learning process in remembering gestures.

\subsection{Potential Techniques} \label{sect:design:techniques}
\begin{figure}[t]
\centering
\includegraphics[width=\linewidth]{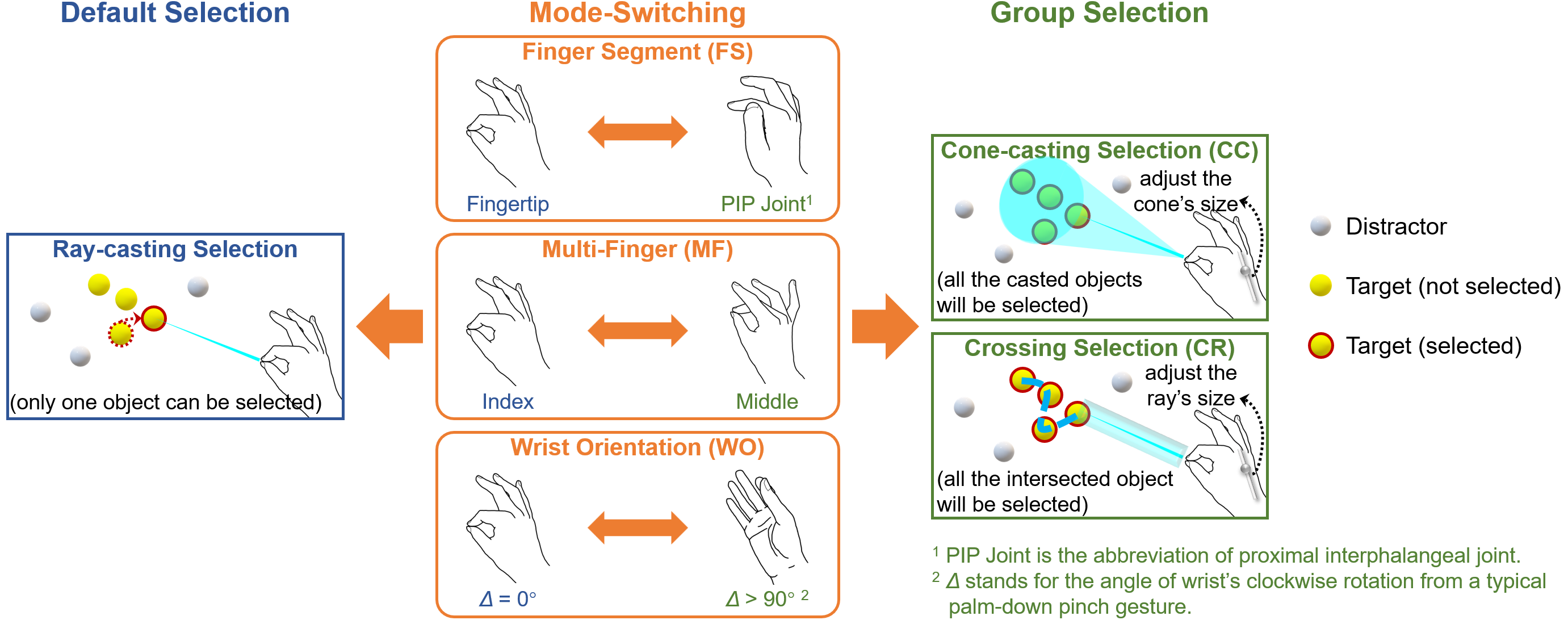}
\caption{\label{fig:technique} The freehand techniques in each part of the multi-object selection process: a ray-casting technique is used for the default selection, Cone-casting Selection or Crossing Selection can do group selection, and the two modes can be transited using Finger Segment, Multi-Finger, or Wrist Orientation gestures. }
\Description{This figure illustrates the proposed techniques for freehand multi-object selection in this work, including a ray-casting technique (via thumb-to-index pinch) for default selection, three mode switching gestures, and two group selection techniques. For mode-switching gestures, Finger Segment (FS) triggers another mode by using the thumb to tap the PIP joint, Multi-Finger (MF) is by thumb-to-middle pinch, and Wrist Orientation (WO) switches the mode via rotating the wrist clockwise for more than 90 degrees. Regarding group selection techniques, Cone-casting Selection (CC) changes the ray to a cone to cover more objects, while Crossing Selection (CR) allows for continuous intersection selection within one selection action.}
\end{figure}

Based on the considerations and the analysis of the interaction process, we derived three parts for multi-object selection: default selection, group selection, and mode switching. As shown in Section~\ref{sect:DG:selection} (Design Goal 1), a ray-casting technique is the most widely used pointing selection technique in VR HMDs. Thus, we use it for default selection. Following the design goals, we propose potential mode-switching and group selection techniques in the next sections. The whole design is illustrated in Figure~\ref{fig:technique}.

\subsubsection{Mode-Switching Techniques}
The default selection is achieved via a freehand ray-casting technique, wherein bringing the dominant hand's fingertips of the thumb and index finger together with the palm facing down (a standard pinch gesture) confirms the selection. We made small adjustments to this pinch gesture to enable a smooth transition to the group selection for our Design Goal 3 (see Section~\ref{sect:DG:MS}). In addition, we concentrated on one-handed mode-switching gestures to reduce the occupation of the non-dominant hand, which may introduce extra fatigue or be used for other interactions. 

\begin{itemize}
    \item \textbf{Finger Segment (FS)}: Instead of the fingertip of the index finger, the user uses her thumb to tap on the proximal interphalangeal (PIP) joint of the index finger to confirm a group selection. 
    \item \textbf{Multi-Finger (MF)}: The user pinches her thumb and middle finger to trigger group selection. 
    \item \textbf{Wrist Orientation (WO)}: When the user's wrist is rotated clockwise from her perspective for more than 90$^\circ$, group selection will be executed if the pinch gesture is performed. 
\end{itemize}

\subsubsection{Group Selection Techniques}
\begin{itemize}
    \item \textbf{Cone-casting Selection (CC)}: The user controls a cone emitted from hand for group selection. Once the user confirms the selection, all the projected objects are selected (or deselected if one was under selection previously). The user can adjust the size of the cone using their non-dominant index finger to drag the slider positioned on their dominant hand. The cone becomes a ray when adjusting the cone to its minimal size (the top of the slider or the closest to the finger). The selection process is discrete; that is, the user needs to release the trigger and do it again for another selection/deselection. 
    \item \textbf{Crossing Selection (CR)}: Crossing selection is a continuous selection process. The user holds the group selection trigger, moves the ray to intersect the target for selection, and releases the trigger for confirmation. Deselection happens when the ray intersects a selected object and is allowed within one selection action. In this work, we leveraged the collision of the ray and the object as the criterion of selection. We also enable the user to adjust the size of the ray in the same way as described in CC. When the user increases the size, the ray looks like a cylinder and becomes easier to collide with the object. 
    \item Rectangle Selection (discarded): The user holds the group selection trigger and formulates a rectangular region similar to the rectangle selection in a desktop interface. The rectangle is instantiated and remains in the X-Y plane where the user starts to draw the rectangle. The user's hand movement in the Z-axis during the group selection is projected to that X-Y plane. When the user releases the trigger, a perspective projection is cast from the user's head position (the origin) to the rectangle and projects to farther planes. All the projected objects are selected/deselected. However, during the pilot testing, we found this 3D viewing process may confuse the users. Furthermore, it was not easy to integrate serial selection and to track accurately using this technique (sometimes the hand moves out of the headset's accurate tracking area). Thus, we discarded this technique from the evaluation. 
    \item Volume Selection (discarded): With Volume Selection, the user points to an object and sends out a pre-defined 3D spherical volume with the pointed object as the center. All the objects within this volume would be selected/deselected. Like CC and CR, the user slides on the dominant hand to adjust the volume's size. We also visualize a replica of the volume (visually equal-sized) above the user's hand to assist her in estimating the target area. However, as the user can only decide the center and define the volume's size by expanding or shrinking it from its center, it is hard to have targets in the volume from all directions accurately. Due to this, Volume Selection was significantly more inefficient than others in our pilot test and was not included in our formal comparisons.  
\end{itemize}

\subsection{Summary} \label{sect:design:summary}
The refined interaction process and selected techniques are visualized in Figure~\ref{fig:technique}. We also illustrate the initial design of the interaction process and techniques in the appendix for reference (Figure~\ref{fig:initialdesign}). We tried to optimize the parameters within each technique through informal testing. Although enhancing them by adding exclusive features for each specific technique is feasible, we focused more on experimental analysis of these techniques with shared features for controlled comparisons. 

Finally, six combinations of mode-switching and group selection techniques have been selected for evaluation. We aimed to investigate user performance and experience of these potential techniques in multi-object selection scenarios in VR. To understand how potential techniques could be incorporated with default selection to provide a smooth workflow, we varied two independent variables: \textsc{MSTech} and \textsc{GSTech}. \textsc{MSTech} represents mode-switching techniques (FS, MF, and WO) while \textsc{GSTech} represents group selection techniques (CC and CR). In the following sections, we call their combinations FS+CC, FS+CR, MF+CC, MF+CR, WO+CC, and WO+CR. 

\section{User Study} \label{sect:exp}
In this user study, we compared the potential techniques in a controlled, simplified test environment with two task complexities (Section~\ref{sect:expt:task}). We followed the guidelines outlined by Bergström et al.~\cite{Bergstrom2021How} to design and report this object selection study.

\subsection{Participants}
We recruited eighteen participants (5 women and 13 men) aged between 19 and 26 years ($M=23$, $SD=2.376$). All of them are right-handed. They have either normal vision ($N=3$) or corrected-to-normal vision ($N=15$). None had claimed they could not see the test environment clearly in the experiment. Sixteen participants reported they were familiar or very familiar with VR/AR/MR HMDs. Ten identified themselves as being familiar or very familiar with mid-air interaction, while two identified their unfamiliarity. 

\subsection{Apparatus}
The study used a Meta Quest Pro VR HMD. Quest Pro has a 106$^\circ$ horizontal field-of-view, an 1800$\times$1920 per eye resolution, and a 90Hz refresh rate. Its inside-out cameras enable 6 degrees of freedom hand tracking. It was connected to a high-performance desktop computer to run the experimental program. The computer was equipped with a Windows 11 system, an Intel Core i9-11900K processor, an NVIDIA GeForce RTX 3090 GPU, and 64GB of RAM. The program was implemented using C\# in Unity (version 2022.3.0f1) with Oculus XR Plugin (version 4.0.0). 

During the experiment, participants sat in front of a table to complete the experiment to minimize fatigue. The experimenter can observe participants' actions in the test environments (the Game view in the Unity interface) through the computer monitor. Figure~\ref{fig:task}(A) illustrates this setup. 

\subsection{Test Environment and Task} \label{sect:expt:task}

\begin{figure}[h]
\centering
\includegraphics[width=\linewidth]{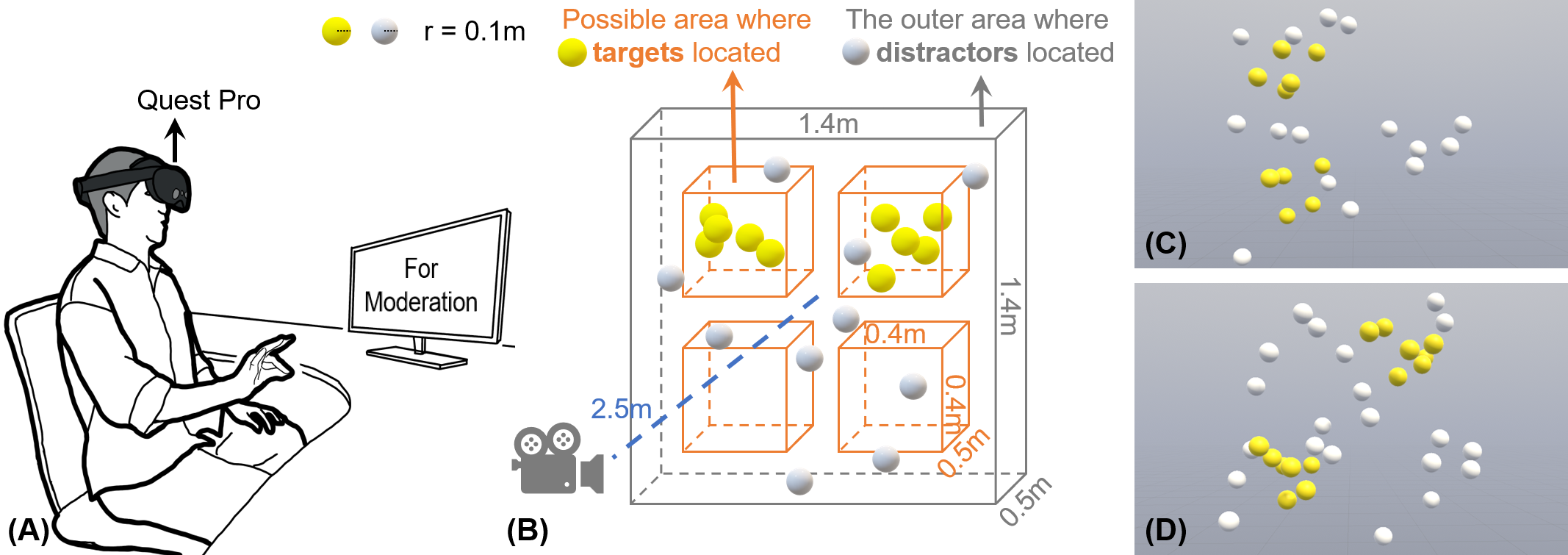}
\caption{\label{fig:task}Illustrations for (A) experimental setup, (B) experimental task, (C) the Low Complexity condition, (D) the High Complexity condition.}
\Description{This figure includes four subfigures to demonstrate our experiments. In sub-figure (A), a participant is completing the task demonstrated via the Quest Pro worn on the head. Sub-figure (B) explains how. Sub-figures (C) and (D) are two screenshots for the Low Complexity condition (5 targets, 10 distractors) and the High Complexity condition (8 targets, 16 distractors), respectively. }
\end{figure}

We used randomized scenarios as test environments to cover more general use cases while applying a few constraints to ensure the given task was controlled to meet our research goal. There were two types of spherical objects in the test environments: \textit{targets} in yellow and \textit{distractors} in grey. They have the same size with a radius of 0.1m. They were all located within a cuboid area of 1.4m$\times$1.4m$\times$0.5m, which was 2.5m in front of a participant's vision. Distractors were randomly positioned in this area. To simulate target groups for group selection, we defined four 0.4m$\times$0.4m$\times$0.5m areas within the outer rectangular area and let targets be generated randomly within two out of these four areas. Figure~\ref{fig:task}(B) demonstrates this task setting. Objects could not be manipulated (like translation or rotation) and only had two states, either selected or not being selected. When an object had been selected, its outline would turn red. Participants could deselect an object by selecting it again, and the deselected object's outline would disappear. 

The test environment involved two levels of task complexity. In the \textit{Low Complexity} condition, 5 targets were randomly generated in each of the target areas (10 targets in total), and 10 distractors were randomly placed in the outer region, as shown in Figure~\ref{fig:task}(C). In the \textit{High Complexity} condition, 8 targets were in each of the target areas (16 targets in total), and 16 distractors were in the outer region, as shown in Figure~\ref{fig:task}(D). That said, the task complexity was controlled by setting the number of objects in the test environment. The Low Complexity condition could be considered as testing in a sparse environment with fewer targets and distractors, and the High Complexity condition was a dense environment with more targets and distractors but kept the same ratio of targets to distractors (1:1). 

Participants were asked to acquire all the targets and avoid selecting distractors as accurately and fast as possible. We explicitly mentioned to them that the priority was accuracy over speed. Typically, participants first performed a group selection and then refined their selection (deselect the distractors or select the missed targets).

\subsection{Experimental Design}
We used a $3\times2$ within-subjects design with \textsc{MSTech} (FS, MF, and WO) and \textsc{GSTech} (CC and CR) as two independent variables, as described in Section~\ref{sect:design:summary}. To minimize the carry-over effect, the order of \textsc{MSTech} $\times$ \textsc{GSTech} conditions was counterbalanced via a balanced Latin Square approach~\cite{James1958Complete}. Within each condition, participants completed ten randomized formal trials, five for each task complexity. Thus, we collected 18 participants $\times$ 3 mode-switching techniques $\times$ 2 group selection techniques $\times$ 2 task complexities $\times$ 5 repetitions = 1080 data trials in total. 

\subsection{Procedure}
Participants first completed a demographic questionnaire and were introduced to the study purpose, design, VR device, and tasks. They were also briefed about the techniques and their controls. Participants then went through the conditions following the given order. Each condition could be divided into three phases. First, participants received a training session to familiarize themselves with the technique. In the training session, they were asked to get used to the given technique in the same task setting as described above. The experimenter explained the technique to participants and then guided them to try all possible controls, including default selection, group selection, deselection, and size adjustment of the ray/cone. The training session included five trials and lasted at least one minute. Second, they completed the ten formal trials. The formal trials were given in a discrete form, where participants needed to click on the button above the object area via a default selection to continue with the next trial. Participants were informed explicitly to complete the ten formal trials in a condition carefully and continuously without rest. Third, they completed questionnaires about their feelings using the given technique (more details in the following section). A short break was given between two conditions. Once participants completed all conditions, they received a semi-structured interview to collect their feedback. The experiment lasted approximately 50 minutes for each participant.

\subsection{Evaluation Metrics}
We have a set of dependent variables involving both objective and subjective measurements. 

\subsubsection{Objective Measurements}
For the objective measurements listed below, we used the average results per condition and participant for statistical analysis. 

\begin{itemize}
    \item \textit{Completion Time}: We recorded the time (in seconds) taken to complete each trial.  
    \item \textit{Number of Errors}: We analyzed the number of missed targets, selected distractors, and total errors (the sum of the prior two). 
    \item \textit{Number of Actions}: The number of actions performed to complete the task. The actions counted included default selection, group selection, and ray/cone adjustment. We counted these actions once they were triggered (i.e., as a discrete action), regardless of how long they have been maintained. 
    \item \textit{Hand Movements}: The total hand movements in meters performed in each trial. It was calculated by aggregating the hand movement distance made in each frame. 
\end{itemize}

\subsubsection{Subjective Measurements}
We also compared the techniques based on subjective measurements, including perceived workload, usability, arm fatigue, and preference rankings.
\begin{itemize}
    \item \textit{NASA-Task Load Index (NASA-TLX)}~\cite{NASA}: A raw NASA-TLX questionnaire was used to measure subjective workload when using the proposed techniques to complete the given task in terms of six dimensions: \textit{mental demand}, \textit{physical demand}, \textit{temporal demand}, \textit{performance}, \textit{effort}, and \textit{frustration}. Further, these six scales derived a weighted \textit{overall score}. 
    \item \textit{System Usability Scale (SUS)}~\cite{SUS}: We used a positive version of the SUS questionnaire to measure the usability of the proposed techniques. The ratings from 10 items were converted to an overall \textit{SUS score} for statistical analysis. 
    \item \textit{Borg CR10 Scale}~\cite{Borg}: \textit{Borg CR10 scale} is a categorical rating with scores ranging from 0 to 10 and corresponding verbal descriptions for assessing perceived arm exertion/fatigue. 
    \item \textit{Ranking}: At the end of the experiment, participants were asked to rank all six techniques according to their overall preference. 
\end{itemize}

Except for the questionnaire for overall preference ranking, all other measurements were collected after participants completed a condition. The performance of the technique in both complexities was taken into consideration. We also interviewed the participants at the end of the experiment. We asked participants to reflect on their experience and share their opinions about the strengths and weaknesses or any other comments about the techniques. 

\subsection{Hypotheses}
Based on our design process and pilot trials, we formulated the following two hypotheses that we were particularly interested in testing in the study: 
\begin{itemize}
    \item \textbf{H1.} Regarding the group selection techniques, CR will outperform CC and will provide a better experience because it involves a continuous selection mechanism, reducing the effort to trigger repeatedly the selection for multiple objects in CC. 
    \item \textbf{H2.} The mode-switching gestures (FS, MF, and WO) will not lead to significantly varying performances and preferences because they are small and easy-to-perform gestures modified from the thumb-to-index pinch gesture.
\end{itemize}

\section{Results} \label{sect:results}

\subsection{Objective Results}
We removed trials in which the completion time was over three standard deviations from the mean ($>M+3SD$) in each condition (15 trials, or 1.39\% of total trials), the number of default selections was more than four times (9 trials, 0.83\%), or the participants skipped by mistake (1 trial, 0.09\%). In total, we removed 25 trials (2.31\%). These trials were treated as outliers and removed because they implied unusual completion of trials (e.g., due to the hand tracking issue). We checked the normality of the data using both Shapiro-Wilk tests and QQ plots. The completion time and number of actions were normally distributed. We then performed repeated-measure (RM-) ANOVA to analyze the effects of the variables and conducted pairwise comparisons with Bonferroni adjustment to the $p$ values. The number of errors (including the number of selected distractors, missed targets, and total errors) and hand movements were not normally distributed. We transformed them via aligned rank transform (ART)~\cite{ART1} before conducting RM-ANOVA tests and applied the ART-C procedure~\cite{ART2} for post-hoc analysis ($p$ values are also Bonferroni-adjusted).

\begin{figure}[h]
  \centering
  \includegraphics[width=\linewidth]{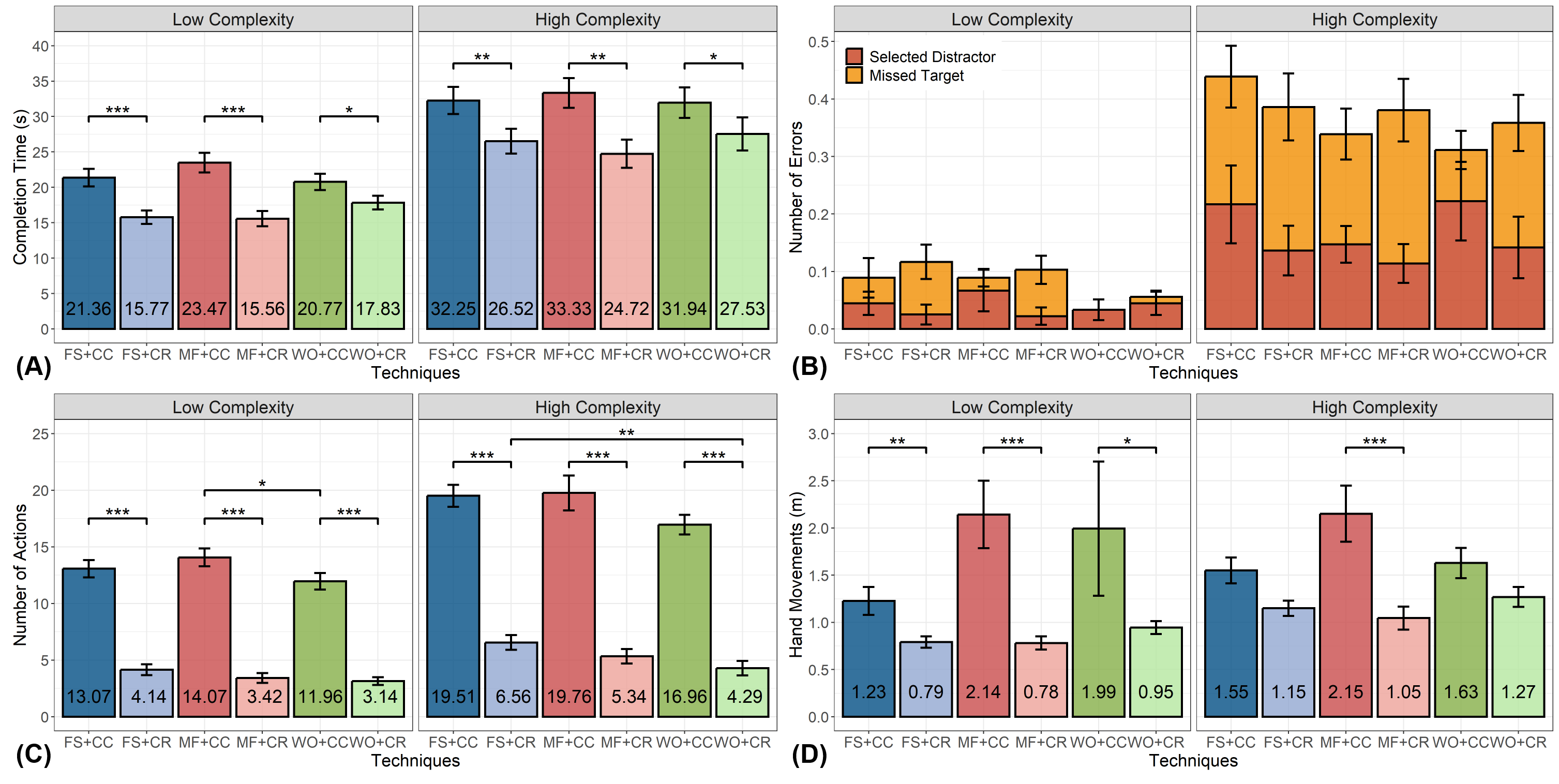}
  \caption{\label{fig:objective} Plots of average ($\pm1SE$) performance under two task complexities. (A) Completion time. (B) Number of errors. (C) Number of actions. (D) Hand movements.}
  \Description{This figure shows the plots of objective results, including (A) completion time, (B) number of errors (number of selected distractors, missed target, and total errors), (C) number of actions, and (D) hand movements.}
\end{figure}

\subsubsection{Completion Time}
In the Low Complexity condition, there was a significant effect of \textsc{GSTech} ($F_{1,17}=40.394, p<0.001, \eta_p^2=0.253$) and a significant interaction effect between \textsc{MSTech} and \textsc{GSTech} ($F_{2,34}=4.528, p=0.018, \eta_p^2=0.045$) on the mean completion time. \textsc{MSTech} did not have a significant effect ($F_{2,34}=0.671, p=0.518, \eta_p^2=0.007$). Pairwise comparisons showed that the mean completion time was significantly longer with CC than CR when the mode-switching technique was FS ($p<0.001$), MF ($p<0.001$), and WO ($p=0.036$). 

In the High Complexity condition, there was a significant effect of \textsc{GSTech} ($F_{1,17}=20.105, p<0.001, \eta_p^2=0.120$) on mean completion time, while the effect of \textsc{MSTech} ($F_{2,34}=0.102, p=0.903, \eta_p^2=0.001$) and the interaction effect ($F_{2,34}=1.728, p=0.193, \eta_p^2=0.011$) were not significant. Similar to the Low Complexity condition, pairwise comparisons showed that CC had a longer completion time than CR when the mode-switching technique was FS ($p=0.003$), MF ($p=0.003$), and WO ($p=0.011$). The results are illustrated in Figure~\ref{fig:objective}(A). 

\subsubsection{Number of Errors}
RM-ANOVA showed that the number of missed targets in the Low Complexity condition was significantly influenced by \textsc{MSTech} ($F_{2,85}=6.424, p=0.003, \eta_p^2=0.131$) and \textsc{GSTech} ($F_{1,85}=5.736, p=0.019, \eta_p^2=0.063$). On the other hand, there was a significant effect of \textsc{GSTech} ($F_{1,85}=5.888, p=0.017, \eta_p^2=0.065$) on the number of missed targets in the High Complexity condition. Except for this, RM-ANOVA did not indicate any other other significant effects. Pairwise comparisons did not show any significant differences among the techniques either. Figure~\ref{fig:objective}(B) visualizes the results. As can be seen, the number of errors was very low (less than one time), regardless of the techniques.

\subsubsection{Number of Actions}
In the Low Complexity condition, there was a significant effect of \textsc{MSTech} ($F_{2,34}=3.586, p=0.039, \eta_p^2=0.042$) and a significant effect of \textsc{GSTech} ($F_{1,17}=229.268, p<0.001, \eta_p^2=0.777$) on the number of actions. The interaction effect between \textsc{MSTech} and \textsc{GSTech} on the number of actions was not significant ($F_{2,34}=2.221, p=0.124, \eta_p^2=0.027$). Pairwise comparisons showed that CC required significantly more actions to complete the task than CR when the mode-switching technique was FS, MF, and WO (all $p<0.001$). Additionally, within the CC group selection technique, MF required significantly more actions than WO ($p=0.022$).

There was a significant effect of \textsc{MSTech} ($F_{2,34}=5.155, p=0.011, \eta_p^2=0.067$) and a significant effect of \textsc{GSTech} ($F_{1,17}=305.758, p<0.001, \eta_p^2=0.748$) on the number of actions in the High Complexity condition. The interaction effect between \textsc{MSTech} and \textsc{GSTech} was not significant ($F_{2,34}=0.795, p=0.460, \eta_p^2=0.010$). Same as in the Low Complexity condition, FS+CC, MF+CC, and WO+CC involved a significantly higher number of actions than FS+CR, MF+CR, and WO+CR, respectively (all $p<0.001$). Furthermore, we also found that FS+CR required significantly more actions than WO+CR ($p=0.003$). Figure~\ref{fig:objective}(C) summarizes the results regarding the number of actions. 

\subsubsection{Hand Movements}
Figure~\ref{fig:objective}(D) shows the results regarding the hand movements. 

RM-ANOVA showed that the hand movements in the Low Complexity condition were significantly influenced by \textsc{MSTech} ($F_{2,85}=4.084, p=0.020, \eta_p^2=0.088$), \textsc{GSTech} ($F_{1,85}=43.879, p<0.001, \eta_p^2=0.340$), and their interaction ($F_{2,85}=5.101, p=0.008, \eta_p^2=0.107$). Results from pairwise comparisons showed that FS+CC, MF+CC, and WO+CC involved a significantly more hand movements than FS+CR ($p=0.004$), MF+CR ($p<0.001$), and WO+CR ($p=0.024$), respectively.

In the High Complexity condition, there was a significant effect of \textsc{GSTech} ($F_{1,85}=44.501, p<0.001, \eta_p^2=0.344$) and a significant interaction effect ($F_{2,85}=4.519, p=0.014, \eta_p^2=0.096$) on the hand movements, while there was no significant effect of \textsc{MSTech} ($F_{2,85}=0.626, p=0.537, \eta_p^2=0.015$). Pairwise comparisons only showed a significant difference between MF+CC and MF+CR ($p<0.001$), with MF+CC having significantly more hand movements.

\subsection{Subjective Results}
\begin{figure}[h]
  \centering
  \includegraphics[width=\linewidth]{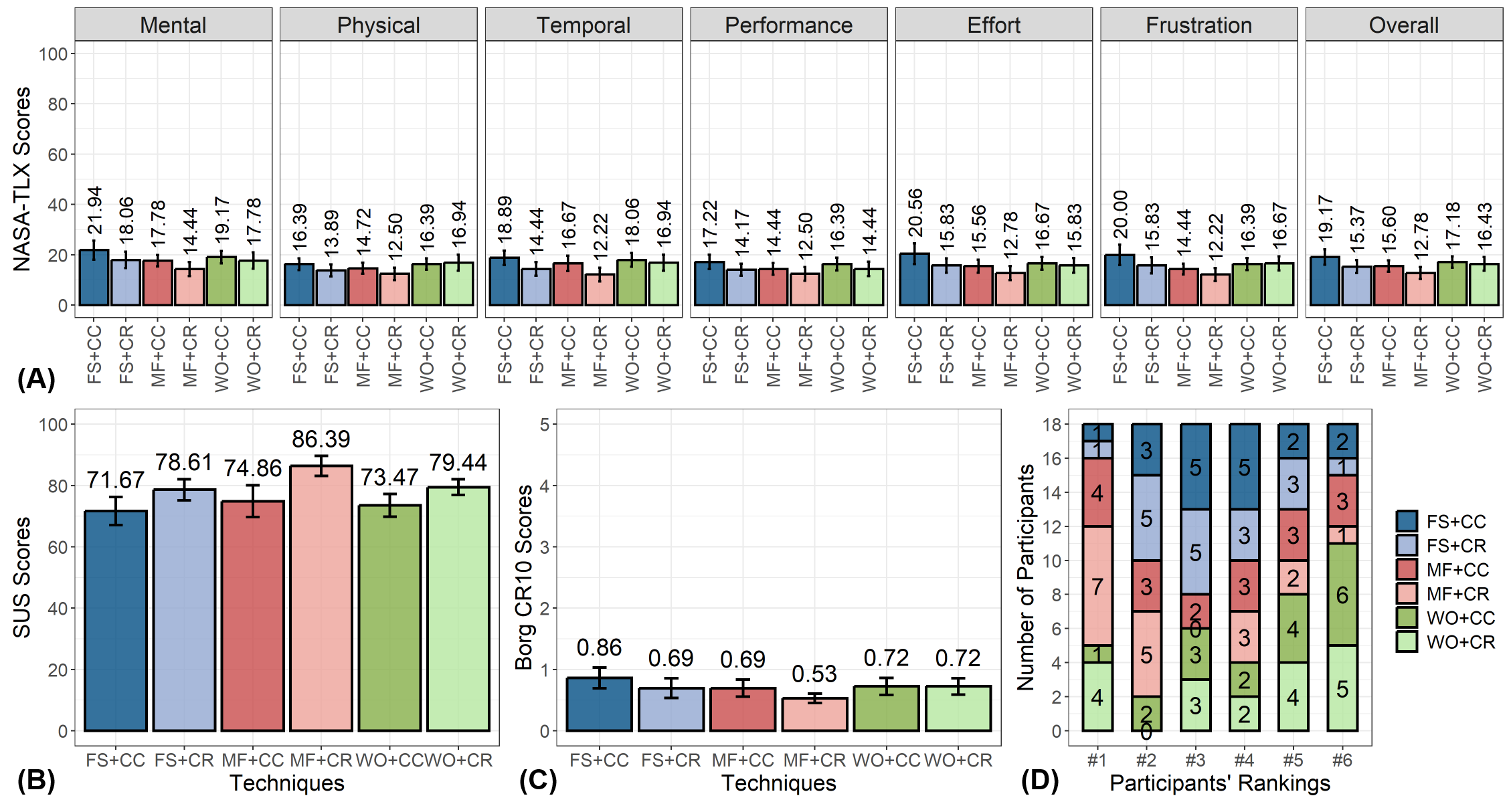}
  \caption{\label{fig:subjective} (A) Average ($\pm1SE$) NASA scores of the techniques. The lower the score is, the lower the perceived workload of the technique (i.e., the better). (B) Average ($\pm1SE$) SUS scores of the techniques. The higher the score is, the higher the usability of the technique (i.e., the better). (C) Average ($\pm1SE$) Borg CR10 scores of the techniques. The lower the score is, the lower the perceived arm fatigue (i.e., the better). (D) Participants' ranking of each technique. }
  \Description{This figure shows the plots of subjective results, including (A) NASA scores, (B) SUS scores, (C) Borg CR10 scores, and (D) participants' ranking. }
\end{figure}

We performed RM-ANOVA and pairwise comparison with Bonferroni adjustments to ART-transformed~\cite{ART1, ART2} questionnaire results, including NASA-TLX scores, SUS scores, and Borg CR 10 scores. The descriptive analyses of these measurements are visualized in Figure~\ref{fig:subjective}(A-C).

Regarding perceived workload, RM-ANOVA revealed significant effects of \textsc{GSTech} on mental demand ($F_{1,85}=4.485, p=0.037, \eta_p^2=0.050$), temporal demand ($F_{1,85}=5.161, p=0.026, \eta_p^2=0.057$), performance ($F_{1,85}=4.058, p=0.047, \eta_p^2=0.046$), and overall score ($F_{1,85}=4.082, p=0.046, \eta_p^2=0.046$). No other significant effect, interaction effect, or post hoc differences were found. As shown in Figure~\ref{fig:subjective}(A), the workload of using each technique to complete the selection task is perceived to be low. 

Both \textsc{MSTech} ($F_{2,85}=5.167, p=0.008, \eta_p^2=0.108$) and \textsc{GSTech} ($F_{1,85}=12.640, p<0.001, \eta_p^2=0.129$) had a significant effect on SUS scores. However, we did not find a significant interaction effect between \textsc{MSTech} and \textsc{GSTech} ($F_{2,85}=0.846, p=0.433, \eta_p^2=0.020$) on SUS scores. Moreover, results from pairwise comparisons showed no significant differences among the techniques. Overall, the proposed techniques were rated with high usability. The average SUS scores for each technique are over 70 points (see Figure~\ref{fig:subjective}(B)).

In terms of Borg CR 10 scores, no significant differences were found. All the techniques were rated with low arm fatigue to complete the multi-object selection tasks, as Figure~\ref{fig:subjective}(C) shows. 

Figure~\ref{fig:subjective}(D) shows participants' ranking of the six techniques. 7 participants (38.89\%) ranked MF+CR the most favored technique, followed by MF+CC ($N=4$, 22.22\%) and WO+CR ($N=4$, 22.22\%). In terms of \textsc{MSTech} (FS vs. MF vs. WO), the figure shows a clear tendency to favor MF-based techniques. Twelve participants (66.67\%) ranked them as the most favored technique. In contrast, WO-based techniques were the least favored because 11 participants (61.11\%) ranked them the last. On the other hand, if we group the ranking according to \textsc{GSTech} (CC vs. CR), participants' preferences were scattered. CR-based techniques were ranked first place by 12 participants (66.67\%) and second place by 10 participants (55.56\%). They were also ranked fifth place by 9 participants (50\%), and sixth place by 7 participants (38.89\%). We present and discuss the interview responses in the Discussion section. 

\section{Discussion} \label{sect:discussion}
Our results demonstrate how the proposed mode-switching techniques and group selection techniques fit into a freehand multi-object selection workflow in randomized VR scenarios. In this section, we discuss the results and evaluation of the techniques and provide design implications that can help the design and development of object selection techniques in VR in the future. 

\subsection{Technique Evaluation}
\textbf{H1} was about the performance and experience of group selection techniques. It was partially supported---the results supported our assumption regarding group selection techniques' performance but contradicted our expectations of their user experience. Many significant performance differences were observed when comparing the two group selection techniques and they showed consistent patterns. CR was faster than CC in completing the multi-object selection tasks in randomized scenarios, regardless of the incorporated mode-switching gesture or the task complexity. In addition, it required fewer actions than CC to complete the task. This is an expected result because CR involves a continuous selection procedure so that users can refine their selection within one selection event. Furthermore, we found participants rarely resized the ray when using CR during the experiments, which also reduced the number of actions. Although CR led to a better performance, some participants raised negative comments on it. They described CR as ``difficult to control'' (P1, P6) or ``unstable'' (P4), especially when the target was surrounded by distractors. P1, P4, and P13 reported that they felt CR was more prone to select an unwanted object, while P13 also mentioned such an error was easy to fix. This might be the reason for participants not to increase the ray size. Though CC required more time and more actions to complete the task, it was preferred by some participants. The primary driver was its selection mechanism---it was consistent with the default pointing selection. P4 and P9 mentioned that they did not need to hold the pinch gesture, which was more relaxed compared to using CR. We also found that CC needed more hand movements in the Low Complexity conditions, which is out of our expectations. We speculate that participants barely adjusted the ray's size when using CR but had to do so with CC. As the ray-casting-based technique, either by crossing or pointing, did not require a large interaction space, the adjustment action involved a greater range of motion by contrast. 

In summary, although CR outperformed CC in the speed, number of actions, and hand movements, CR and CC both led to a good user experience and preferred by participants to be integrated into the freehand multi-object selection workflow. The ray adjustment feature may not needed for CR in randomized scenarios as it behaves sensibly for some users. On the other hand, to further improve the CC, four participants (P4, P6, P11, P18) suggested giving visual cues when the objects are illuminated before the selection, such as showing them with an outline of another color. P11 also suggested allowing users to customize the direction of the slider.

\textbf{H2} regarding the performances and preferences in mode-switching gestures was also partially supported. We did not observe much significant performance differences when comparing the mode-switching gestures. These gestures only have minimal modifications from the standard pinch gestures, and did not impact the performance much. However, the participants' ranking and their feedback revealed that they have different preferences toward these gestures (which contradicted to \textbf{H2}). Overall, MF is the most favored mode-switching gesture and was acceptable to most participants (see Figure~\ref{fig:subjective}(D)). On the contrary, most criticism was given to WO. P10 said, ``\textit{when using WO, I pay extra attention to how much has been or still needs to rotate.}'' After switching to the group selection mode, maintaining the gesture rotated was also more challenging (P6, P9, P10). P6 mentioned he felt controlling the ray/cone with palm facing up or close to up (WO) was more difficult and unnatural than with palm facing down (in a standard pinch, FS, or MF). As for FS, P6 and P17 thought this gesture was not natural and not commonly used even in daily life. On the other hand, P1 felt that it was too close to the fingertip pinch (from the perspective of distance and interaction), taking her some time to distinguish between them and remember. 

In this user study, we used randomized testing environments with several constraints to compare the proposed techniques. Based on the results and our observations, completing the High Complexity tasks was clearly tougher than the Low Complexity tasks during the experiments. P1 and P15 expressed their concerns: ``\textit{I think I cannot select the targets accurately anymore if more objects crowded there.}'' When the VR environment becomes more complex, such as having small or occluded objects, a disambiguation technique may be necessary to resolve the ambiguity. Future studies may explore how to insert a suitable disambiguation technique into the workflow to acquire multiple targets with more ease and precision. On the other hand, we forced the targets to be generated in two clusters to investigate the group selection technique. In the actual applications, the targets may be placed in a structured layout, which should make the group selection techniques more effective. Considering the participants' workload, we could not test the techniques in these scenarios in this study, but we mark the importance of this evaluation for future studies.

\subsection{Design Implications}
Based on the study results, we distilled the following three design implications. 

\begin{itemize}
    \item Both Crossing Selection (CR) and Cone-casting Selection (CC) are suitable for group-based selection. If the selection performance is critical to the application scenario, CR is superior to CC. 
    \item Using the middle finger (MF) for switching between the single-object selection and multi-object selection modes is suggested. 
    \item Rotating the wrist (WO) for switching the selection mode is not preferred by users and thus not suggested. 
\end{itemize}

\section{Limitations and Future Work}
There are three limitations to our work. First and foremost, we only compared the techniques in randomized scenarios due to the size of the study. We chose to start with a random arrangement of multiple objects because the derived results and findings can be relatively more universal and generalizable. It is worth evaluating the techniques in more scenarios, including randomized scenarios with varying constraints and different types of structured layouts (see also Section~\ref{sect:discussion}), and collecting more types of user performance for deriving user behavior patterns (e.g., the most frequently used cone's size in Cone-Casting selection). Further, we also want to investigate their use in practical applications involving multi-object selection, such as data visualization, 3D modeling, and building design applications. 
Second, our participants were all right-handed. For future studies, we would like to investigate the performance of left-handed users. We also want to invite more participants with more diverse backgrounds (in terms of gender, age group, past experience, and degrees of motor challenges) to collect their feedback and invite them to use the techniques in the long term to investigate their prolonged use. By doing so, we could also provide further insights into the individual and group differences in the performance and user experiences of the techniques. Third, we focused on the interaction techniques that were applicable to current available HMDs, and due to this, we used their built-in hand-tracking modules. We acknowledged that the precision and stability of hand-tracking may affect the results. In the future, we want to compare the proposed techniques with more micro-interactions that are not usable now but become feasible with the advancements in tracking technologies.

\section{Conclusion}
In this work, we present an analysis of freehand multi-object selection techniques in virtual reality (VR). Specifically, we investigate how different group selection techniques and mode-switching gestures impact the performance and user experience in multi-object selection tasks. With the results from a user study with eighteen participants, we found crossing selection to be fast and required fewer actions and hand movements compared to cone-casting selection, while both techniques led to a good user experience and gained acceptance by participants. For transitioning between the default single-object selection mode and multi-object selection mode, the three proposed mode-switching gestures showed comparable performance. Based on the participants' feedback, using the middle finger pinch gesture was recommended while oriented the pinch gesture was not. We hope these results and findings can be useful to the practitioners of the VR community in designing and developing more usable gestural interactions.


\bibliographystyle{ACM-Reference-Format}
\bibliography{sample-base}

\appendix

\section{Initial Design of Interaction Process and Techniques}

\begin{figure}[h]
\centering
\includegraphics[width=\linewidth]{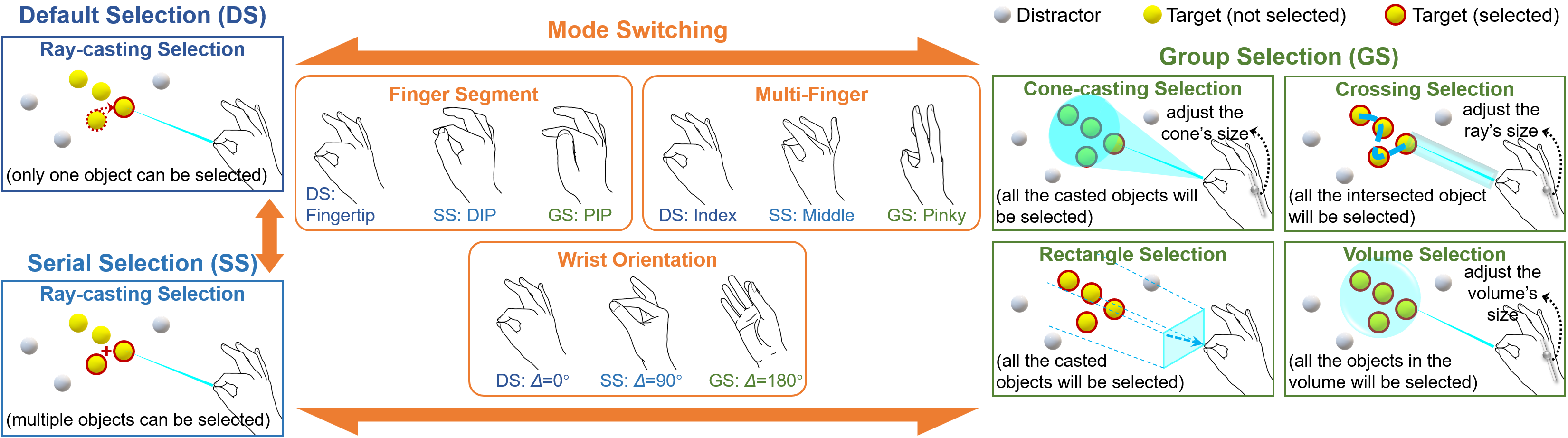}
\caption{\label{fig:initialdesign} The initial design of the multiple-object selection process and freehand techniques. }
\Description{This figure illustrates the initial design of the interaction process and techniques. In this design, the serial selection is not combined with the group selection and requires a unique activation gesture. Mode-switching techniques support the transitions for three modes. In addition, two more group selection techniques, Rectangle Selection and Volume Selection, are included.}
\end{figure}

Figure~\ref{fig:initialdesign} illustrates the initial design of the interaction process and techniques for freehand multi-object selection in VR HMDs. In this version, the serial selection is a separate selection mode with a unique activation gesture, and Rectangle Selection and Volume Selection are included. After pilot tests, we optimized the whole process. More has been discussed in Section~\ref{sect:design}.

\end{document}